\documentclass[twocolumn,showpacs,preprintnumbers,aps,prl,amsmath,amssymb,superscriptaddress]{revtex4}

\usepackage{graphicx}
\usepackage{dcolumn}
\usepackage{bm}
\usepackage{amssymb}
\usepackage[german, english]{babel}
\usepackage{dcolumn} 
\begin{document}
\title{Exploring an ultracold Fermi-Fermi mixture:\\Interspecies Feshbach resonances and scattering properties of $^{6}$Li and $^{40}$K}
\author{E.\ Wille}
 \affiliation{Institut f\"ur Quantenoptik und Quanteninformation,
\"Osterreichische Akademie der Wissenschaften, 6020 Innsbruck,
Austria}
 \affiliation{Institut f\"ur Experimentalphysik und
Forschungszentrum f\"ur Quantenphysik, Universit\"at Innsbruck,
6020 Innsbruck, Austria}
\author{F.M.\ Spiegelhalder}
\author{G.\ Kerner}
\author{D.\ Naik}
\author{A.\ Trenkwalder}
\author{G.\ Hendl}
\author{F.\ Schreck}
 \affiliation{Institut f\"ur Quantenoptik und Quanteninformation,
\"Osterreichische Akademie der Wissenschaften, 6020 Innsbruck,
Austria}
\author{R.~Grimm}
\affiliation{Institut f\"ur Quantenoptik und Quanteninformation,
\"Osterreichische Akademie der Wissenschaften, 6020 Innsbruck,
Austria}
 \affiliation{Institut f\"ur Experimentalphysik und
Forschungszentrum f\"ur Quantenphysik, Universit\"at Innsbruck,
6020 Innsbruck, Austria}
\author{T.G.\ Tiecke}
\author{J.T.M.\ Walraven}
 \affiliation{Van der Waals-Zeeman Institute of the University of Amsterdam, 1018 XE, The
 Netherlands}
\author{S.J.J.M.F.\ Kokkelmans}
 \affiliation{Eindhoven University of Technology, P.O. Box 513, 5600 MB Eindhoven, The
 Netherlands}
\author{E.\ Tiesinga}
\author{P.S.\ Julienne}
 \affiliation{Joint Quantum Institute, National Institute of Standards and Technology and University of Maryland, Gaithersburg, Maryland 20899-8423, USA}

\date{\today}

\pacs{34.50.-s, 67.85.-d, 05.30.Fk}

\begin{abstract}
\textit{Abstract} We report on the observation of Feshbach
resonances in an ultracold mixture of two fermionic species,
$^6$Li and $^{40}$K. The experimental data are interpreted using a
simple asymptotic bound state model and full coupled channels
calculations. This unambiguously assigns the observed resonances
in terms of various $s$- and $p$-wave molecular states and fully
characterizes the ground-state scattering properties in any
combination of spin states.
\end{abstract}

\maketitle

Fermion pairing and Fermi superfluidity are key phenomena in
superconductors, liquid $^3$He, and other fermionic many-body
systems. Our understanding of the underlying mechanisms is far
from being complete, in particular for technologically relevant
high-T$_{\rm c}$ superconductors. The emerging field of ultracold
atomic Fermi gases has opened up unprecedented possibilities to
realize versatile and well-defined model systems. The control of
interactions, offered in a unique way by Feshbach resonances in
ultracold gases, is a particularly important feature. Such
resonances have been used to achieve the formation of bosonic
molecules in Fermi gases and to control pairing in many-body
regimes
\cite{Inguscio2006ufg,Giorgini2007tou,Chin2008fri,Kohler2006poc,Bloch2007mbp}.

So far all experiments on strongly interacting Fermi systems have
been based on two-component spin mixtures of the same fermionic
species, either $^6$Li or $^{40}$K
\cite{Inguscio2006ufg,Giorgini2007tou}. Control of pairing is
achieved via a magnetically tunable $s$-wave interaction between
the two states. After a series of experiments on balanced spin
mixtures with equal populations of the two states, recent
experiments on $^6$Li have introduced spin imbalance as a new
degree of freedom and begun to explore novel superfluid phases
\cite{Zwierlein2006fsw,Partridge2006papetal}. Mixing two different
fermionic species leads to unprecedented versatility and control.
Unequal masses and the different responses to external fields lead
to a large parameter space for experiments and promise a great
variety of new phenomena
\cite{Liu2003igs,Petrov2005dmi,Iskin2006tsf,Orso2007ead,Petrov2007cpoetal}.
The combination of the two fermionic alkali species, $^6$Li and
$^{40}$K, is a prime candidate to realize strongly interacting
Fermi-Fermi systems.

\begin{figure}
\includegraphics[width=84 mm]{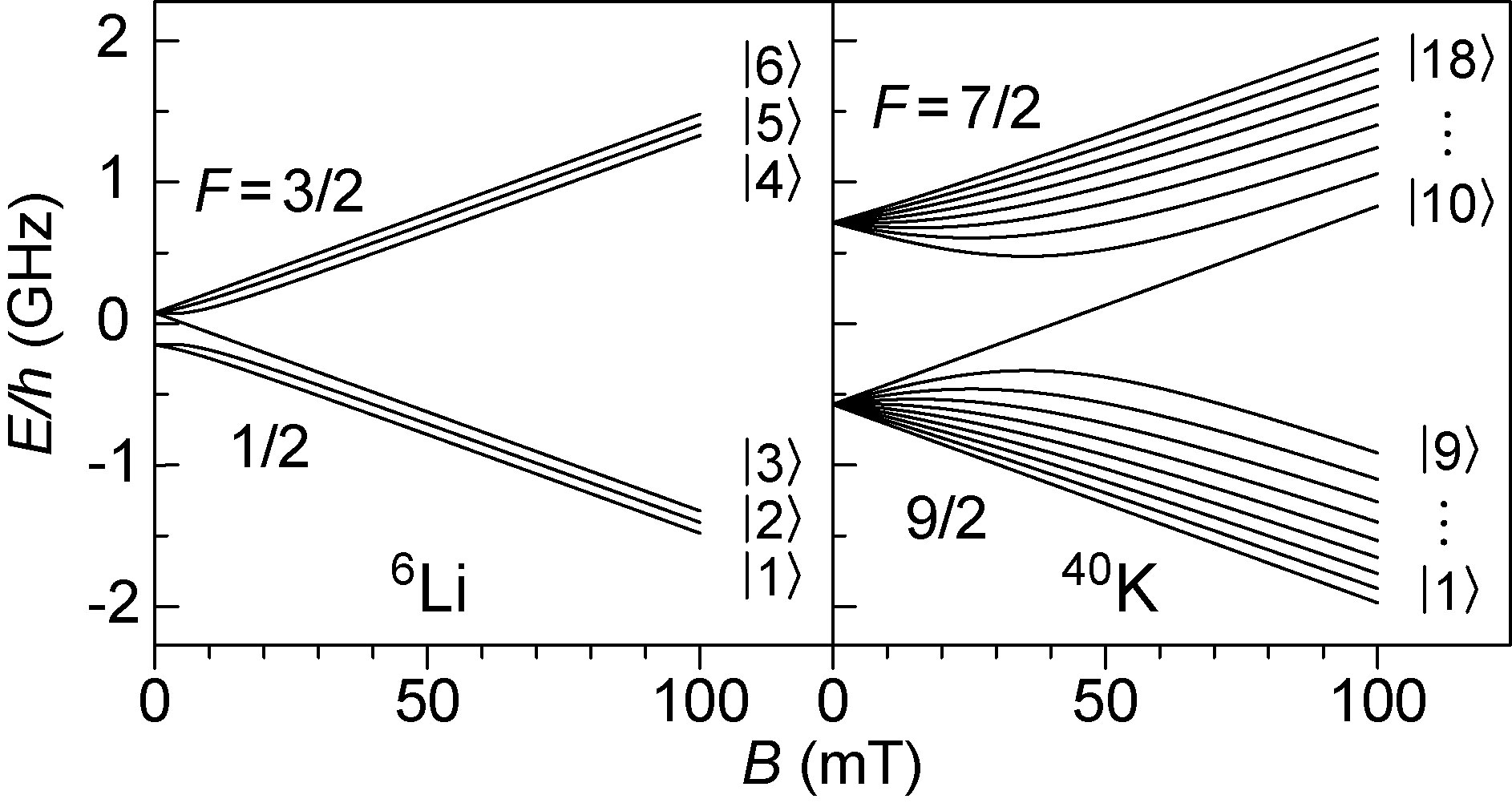}
\caption{\label{fig:LevelScheme} Ground state energies of $^6$Li
and $^{40}$K versus magnetic field.}
\end{figure}

\begin{figure*}
\includegraphics[width=17.6cm]{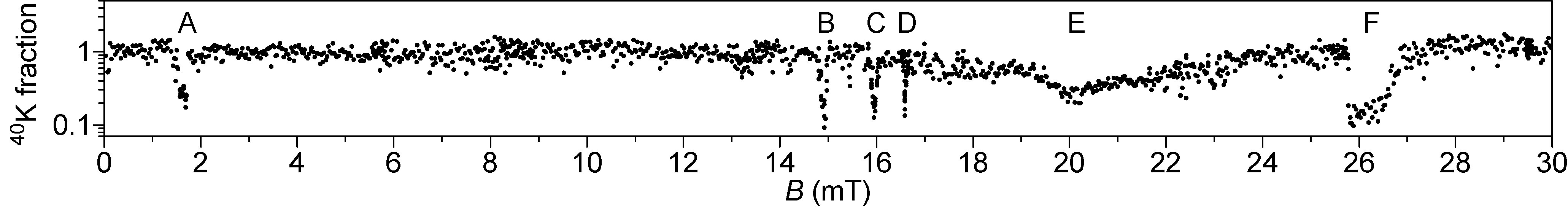}
\caption{\label{fig:Li1K2FeshbachScan} Feshbach scan of the
Li$|1\rangle$K$|2\rangle$ mixture. The remaining fraction of
$^{40}$K atoms relative to off-resonant regions after 10\,s
interaction with $^6$Li atoms is shown as a function of magnetic
field. Loss features A, B, C, D, and F are due to interspecies
Feshbach resonances. Loss feature E is caused by a $^{40}$K
$p$-wave Feshbach resonance \cite{Regal2003tpw}.}
\end{figure*}

In this Letter, we realize a mixture of $^6$Li and $^{40}$K and
identify heteronuclear Feshbach resonances
\cite{Stan2004oofetal,Inouye2004oohetal,Ferlaino2006fsoetalibid}.
This allows us to characterize the basic interaction properties.
Figure~\ref{fig:LevelScheme} shows the atomic ground-state energy
structure. We label the energy levels Li$|i\rangle$ and
K$|j\rangle$, counting the states with rising energy. The
hyperfine splitting of $^6$Li is $(3/2)a^{\rm Li}_{\rm hf}/h =
228.2\,$MHz. For $^{40}$K, the hyperfine structure is inverted and
the splitting amounts to $(9/2)a^{\rm K}_{\rm hf}/h=
-1285.8\,$MHz~\cite{Arimondo1977edo}. For the low-lying states
with $i \le 3$ and $j \le 10$, the projection quantum numbers are
given by $m_{\rm Li} = -i+3/2$ and $m_{\rm K} = j-11/2$. A
Li$|i\rangle$K$|j\rangle$ mixture can undergo rapid decay via spin
relaxation if exoergic two-body processes exist that preserve the
total projection quantum number $M_{\rm F} = m_{\rm Li} + m_{\rm
K} = -i + j -4$. Whenever one of the species is in the absolute
ground state and the other one is in a low-lying state ($i=1$ and
$j\le10$ or $j=1$ and $i\le3$), spin relaxation is strongly
suppressed \cite{Simoni2003mcoetal}.

We prepare the mixture in an optical dipole trap, which is formed
by two 70\,W-laser beams (wavelength 1070\,nm), crossing at an
angle of 12$^{\circ}$~\cite{moredetails}. The dipole trap is
loaded with about $10^7$ $^6$Li atoms and a few $10^4$ $^{40}$K
atoms from a two-species magneto-optical trap (MOT). At this stage
the trap depth for $^6$Li ($^{40}$K) is $1.7\,$mK ($3.6\,$mK) and
the trap oscillation frequencies are $13\,$kHz ($7.3\,$kHz) and
$1.7\,$kHz ($1.0\,$kHz) in radial and axial directions. After
preparation of the internal states of the
atoms~\cite{moredetails}, a balanced mixture of Li$|1\rangle$ and
Li$|2\rangle$ atoms together with K$|1\rangle$ atoms is obtained.
We perform evaporative cooling at a magnetic field of 76\,mT close
to the 83.4\,mT Feshbach resonance between Li$|1\rangle$ and
Li$|2\rangle$ \cite{Inguscio2006ufg,Giorgini2007tou} by reducing
the optical dipole trap depth exponentially by a factor of 70 over
2.5\,s. We observe that potassium remains thermalized with lithium
during the evaporation. This results in $10^5$ Li$|1\rangle$ and
$10^5$ Li$|2\rangle$ atoms together with $10^4$ K$|1\rangle$ atoms
at a temperature of $4\,\mu$K. This three-component Fermi mixture
serves as a starting point to prepare several different stable
two-component mixtures, namely Li$|2\rangle$K$|1\rangle$,
Li$|1\rangle$K$|1\rangle$, Li$|1\rangle$K$|2\rangle$, or
Li$|1\rangle$K$|3\rangle$ with $M_{\rm F}=-5,-4,-3,-2$,
respectively. Atoms in the K$|1\rangle$ state are transferred to
the desired state with adiabatic radio-frequency sweeps.
Population in unwanted states is pushed out of the trap by pulses
of resonant light \cite{moredetails}. Finally, to increase the
collision rate, the sample is compressed by increasing the power
of the optical trap. The temperature rises to 12\,$\mu$K and the
peak density of lithium (potassium) increases to about
$10^{12}$cm$^{-3}$ (few $10^{11}$cm$^{-3}$).

We detect Feshbach resonances by observing enhanced atom loss at
specific values of the magnetic field~\cite{Chin2008fri}, which is
caused by three-body decay. For each mixture we perform a magnetic
field scan with a resolution of 0.03\,mT between 0 and 74\,mT (0
to 40\,mT for the Li$|1\rangle$K$|3\rangle$ mixture). A scan
consists of many experimental cycles, each with a total duration
of about one minute during which the mixture is submitted for ten
seconds to a specific magnetic field value. The quantity of
remaining atoms is measured by recapturing the atoms into the MOTs
and recording their fluorescence light.

In Fig.~\ref{fig:Li1K2FeshbachScan}, we show a loss spectrum of
Li$|1\rangle$K$|2\rangle$. A striking feature is that the
potassium atom number decreases by an order of magnitude at
specific values of the magnetic field. Since the mixture contains
an order of magnitude more lithium than potassium atoms, the
lithium atom number does not change significantly by interspecies
inelastic processes. Therefore, the potassium loss is exponential
and near complete. In order to distinguish loss mechanisms
involving only one species from those involving two species, we
perform additional loss measurements, using samples of either pure
$^6$Li or pure $^{40}$K. Loss features A, B, C, D, and F only
appear using a two-species mixture. Loss feature E persists in a
pure $^{40}$K sample and can be attributed to a potassium $p$-wave
Feshbach resonance~\cite{Regal2003tpw}. On the basis of the
experimental data only, we can not unambiguously attribute loss
feature C to an interspecies Feshbach resonance, since it
coincides with a known $^6$Li $p$-wave
resonance~\cite{Zhang2004pwfetal,Schunck2005frietal}.

Our main findings on positions and widths $\Delta B$ of the
observed loss features are summarized in Table
\ref{tab:PotassiumLossFeatures}, together with the results of two
theoretical models described in the following.

Our analysis of the data requires finding the solutions for the
Hamiltonian $H=H^{\rm hf}_{\rm \alpha}+H^{\rm hf}_{\rm
\beta}+H^{\rm rel}$. To underline the generality of our model, we
refer to Li as $\alpha$ and to K as $\beta$. The first two terms
represent the hyperfine and Zeeman energies of each atom,
 $H^{\rm hf} = (a_{\rm hf}/\hbar^2) \mathbf s \cdot
\mathbf i  + \gamma_e \mathbf s \cdot \mathbf B -\gamma_n \mathbf
i \cdot \mathbf B $, where $\mathbf s$ and $\mathbf i$ are the
single-atom electron and nuclear spin, respectively, and
$\gamma_e$ and $\gamma_n$ are the respective gyromagnetic ratios.
The Hamiltonian of relative motion is
\begin{equation} \label{Hrel}
 H^{\rm rel} = \frac{\hbar^2}{2\mu} \left(- \frac{d^2}{dr^2} + \frac{l(l+1)}{r^2}\right) + \sum_{S=0,1} V_S(r) P_S,
\end{equation}
where $\mu$ is the reduced mass, $r$ is the interatomic
separation, and $l$ is the angular momentum quantum number for the
relative motion. Defining the total electron spin as $\mathbf S =
\mathbf s_\alpha + \mathbf s_\beta$, the projection operator $P_S$
either projects onto the $S=0$  singlet or $S=1$ triplet spin
states.  The potential $V_S(r)$ is thus either for the singlet
X$^1\Sigma$ or triplet a$^3\Sigma$ state. This Hamiltonian $H$
conserves both $l$ and $M_{\rm F}$.

Our first method to locate the Feshbach resonances is inspired by
a two-body bound state model for homonuclear~\cite{Moerdijk95riu}
and heteronuclear~\cite{Stan2004oofetal} systems. We have expanded
this previous work to include the part of the hyperfine
interaction that mixes singlet and triplet levels. This mixing is
crucial for the present analysis. We refer to this model as the
Asymptotic Bound state Model (ABM).

The ABM model expands the bound state solutions $|\Psi^l\rangle $
for each $l$ in terms of $|\psi_S^l \rangle
|S,M_S,\mu_{\alpha},\mu_{\beta}\rangle$ where $|\psi_S^l \rangle$
is the asymptotic last bound eigenstate of the potential
$V_S(r)+\hbar^2 l(l+1)/(2 \mu r^2)$ and
$|S,M_S,\mu_{\alpha},\mu_{\beta}\rangle$ are spin functions where
$M_S$, $\mu_{\alpha}$ and $\mu_{\beta}$ are the magnetic quantum
numbers of $\mathbf S$, $\mathbf i_{\alpha}$ and $\mathbf
i_{\beta}$, respectively. Only spin functions with the same
conserved $M_{\rm F}=M_S+\mu_{\alpha}+\mu_{\beta}$ are allowed.
Note that $S,M_S,\mu_{\alpha},\mu_{\beta}$ are good quantum
numbers for large magnetic field. Expanding $|\Psi^l\rangle $ in
this basis and assuming that the overlap $\langle
\psi_0^l|\psi_1^l\rangle$ is unity~\cite{endnoteOverlap}, the
coupled bound state energies are found by diagonalizing the
interaction matrix~\cite{moredetails}.

\begin{table}
\caption{\label{tab:PotassiumLossFeatures} Feshbach resonances in
collisions between $^6$Li and $^{40}$K in a range from 0 to
76\,mT. For their positions $B_0$, we give the center of the
measured loss features and the results from both the ABM and
coupled channels calculations. The first columns give the $^6$Li
and $^{40}$K channel indices $i$ and $j$ and the projection
quantum number $M_{\rm F} = -i+j-4$. Note that the experimental
width of a loss feature, $\Delta B$, is not the same thing as the
width $\Delta B_{\rm s}$ related to the scattering length
singularity. The latter is only defined for $s$-wave resonances,
and not for the observed $p$-wave resonances. The typical
statistical and systematic error in the experimental $B_0$ is
about 0.05\,mT for $s$-wave resonances. }
\begin{ruledtabular}
\begin{tabular}{ccccccc}
& & \multicolumn{2}{c}{Experiment} & \multicolumn{1}{c}{ABM} & \multicolumn{2}{c}{Coupled channels}\\
$i,\,j$ & $M_{\rm F}$ & $B_0$ & $\Delta B$ & $B_0$  & $B_0$  & $\Delta B_{\rm s}$ \\
 & & (mT) &  (mT) &  (mT) & (mT) & (mT) \\
\hline
2,\,1 & -5 & 21.56$^{a}$ & 0.17 & 21.67 & 21.56 & 0.025\\
1,\,1 & -4 & 15.76 & 0.17 & 15.84 & 15.82 & 0.015\\
1,\,1 & -4 & 16.82 & 0.12 & 16.92 & 16.82 & 0.010\\
1,\,1 & -4 & 24.9 & 1.1 & 24.43 &  24.95 & $p$ wave\\
1,\,2 & -3 & 1.61 & 0.38 & 1.39 & 1.05 & $p$ wave\\
1,\,2 & -3 & 14.92 & 0.12 & 14.97 & 15.02 & 0.028\\
1,\,2 & -3 & 15.95$^{a}$ & 0.17 & 15.95 & 15.96 & 0.045\\
1,\,2 & -3 & 16.59 & 0.06 & 16.68 & 16.59 & 0.0001\\
1,\,2 & -3 & 26.3 & 1.1 & 26.07 & 26.20 & $p$ wave\\
1,\,3 & -2 & \multicolumn{2}{c}{not observed} & 1.75 & 1.35 & $p$ wave\\
1,\,3 & -2 & 14.17 & 0.14 & 14.25 & 14.30 & 0.036\\
1,\,3 & -2 & 15.49 & 0.20 & 15.46 & 15.51 & 0.081\\
1,\,3 & -2 & 16.27 & 0.17 & 16.33 & 16.29 & 0.060\\
1,\,3 & -2 & 27.1 & 1.4 & 27.40 & 27.15 & $p$ wave\\
\end{tabular}
\end{ruledtabular}
\footnotetext[1]{Near coincidences with lithium $p$-wave
resonances~\cite{Zhang2004pwfetal,Schunck2005frietal}.}
\end{table}

The energies $E_S^l$ of the last bound state of the $S=$ 0 and 1
potentials are eigenvalues of Eq.\,(\ref{Hrel}), and serve as free
parameters in the ABM model. We can reduce this to only two
binding energy parameters $E_0=-E_0^0$ and $E_1=-E_1^0$ if we use
information about the actual shape of the potential.  We can do
this using model potentials derived from
Refs.~\cite{Salami2007afaetal} and~\cite{Aymar05}, and the van der
Waals coefficient $C_6=2322\,E_{\rm h}\,a_0^6$ ($E_{\rm
h}=4.35974\times 10^{-18}$ J and
$a_0=0.0529177$\,nm)~\cite{Derevianko01}. Each $E_S$ can be varied
by making small changes to the short range potential while keeping
$C_6$ fixed. The energy $E_S$ uniquely determines both the
$s$-wave scattering length as well as $E_S^l$ for $l>0$.

\begin{figure}
  \includegraphics[width=84mm]{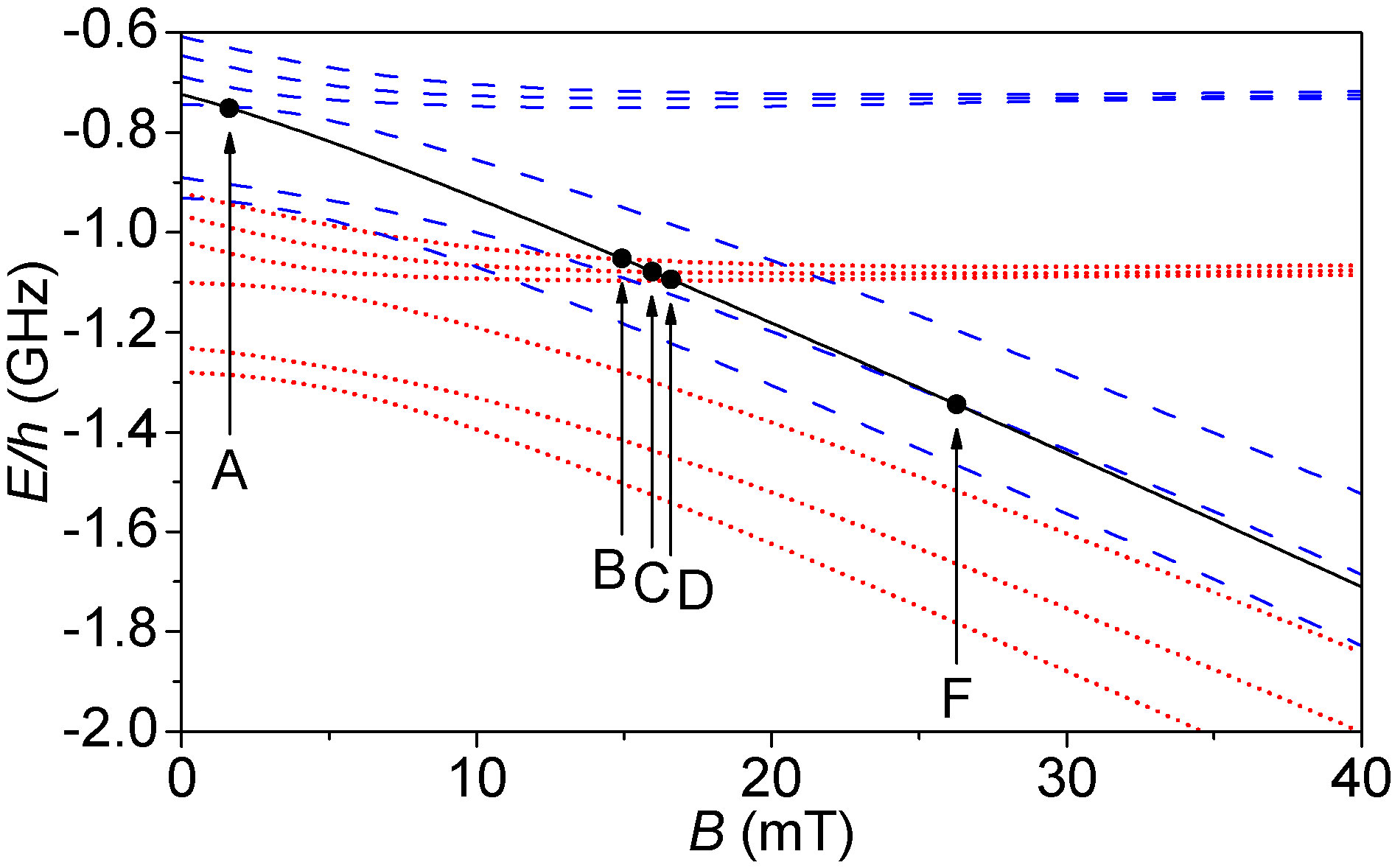}
\caption{Bound state energies versus magnetic field. Dotted
(dashed) lines indicate the $s$-wave ($p$-wave) states. The
two-body threshold for the Li$|1\rangle$K$|2\rangle$ collision
channel ($M_{\rm F}=-3$) is indicated by the solid line. The dots
and the corresponding arrows indicate the measured resonance
positions (see Fig. \ref{fig:Li1K2FeshbachScan}).}
\label{theoryfig1}
\end{figure}

Figure~\ref{theoryfig1} shows the bound state energies of the ABM
model as a function of magnetic field for $M_{\rm F} = -3$.
Feshbach resonances occur at the crossings of bound states and
threshold. We find a good fit for the experimental resonance
positions for parameters $E_0/h$ = 716(15)\,MHz and $E_1/h$ =
425(5)\,MHz, where the uncertainty represents one standard
deviation, see Table \ref{tab:PotassiumLossFeatures}.

\begin{figure}
   \includegraphics[width=84mm]{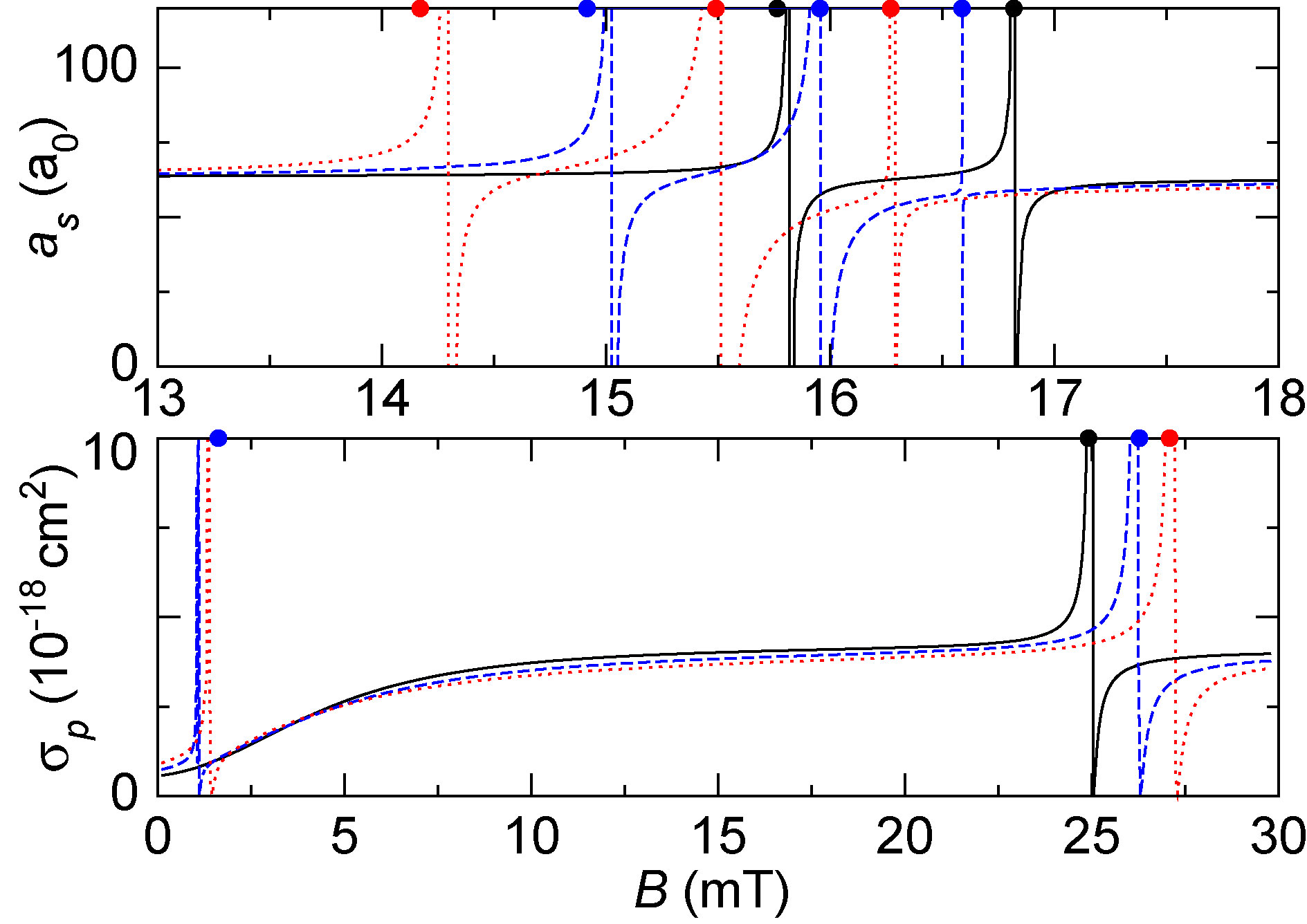}
\caption{Results from coupled channels calculations for the
magnetic-field dependence of the $s$-wave scattering length
$a_s(B)$ (upper panel) and the $m_l=0$ contribution to the
$p$-wave elastic scattering cross section $\sigma_p (E)$  for
$E/k_B= 12\,\mu$K (lower panel) for the channels in Table
\ref{tab:PotassiumLossFeatures} with $M_{\rm F}=-4$ (solid line),
$-3$ (dashed line), and $-2$ (dotted line). The dots indicate the
measured resonance locations.} \label{theoryfig2}
\end{figure}

For additional analysis we have also used exact, yet much more
computationally complex coupled channels
calculations~\cite{Stoof1988sea}, varying the short range
potential as discussed above. An optimized fit to the measured
resonance positions gives $E_0/h$=721(10)\,MHz and $E_1/h$ =
426(3)\,MHz. This corresponds to a singlet scattering length of
52.1(3)\,$a_0$ and a triplet scattering length of 63.5(1)\,$a_0$.
Thus, within the fitting accuracy to the experimental data, the
prediction of the ABM model agrees with the result of the full
coupled channels calculation. Table
\ref{tab:PotassiumLossFeatures} shows the coupled channels
resonance locations and widths for a representative calculation
with $E_0/h$ = 720.76\,MHz and $E_1/h$ = 427.44\,MHz. The $s$-wave
resonance width $\Delta B_{\rm s}$ is defined by $a_{\rm s}(B) =
a_{\rm bg} (1 - \Delta B_{\rm s} /(B - B_0 ))$, where $a_{\rm bg}$
is the background scattering length near the resonance position
$B_0$. Note that $\Delta B_{\rm s}$ need not be the same as the
empirical width $\Delta B$ of a loss-feature. All resonances
except the $M_{\rm F} = -3$ $p$-wave resonance near 1.6\,mT agree
with the measured positions within 0.13\,mT. Fine-tuning of the
long range potential would be needed to fit this resonance to
comparable accuracy. Figure~\ref{theoryfig2} shows the calculated
$s$-wave scattering lengths and $p$-wave elastic cross sections
versus magnetic field $B$ for this model. The background
scattering length $a_\mathrm{bg}$ for the $s$-wave resonances is
approximately 63\,a$_0$.

The accuracy and computational simplicity of the ABM model make
resonance assignments very efficient, allowing rapid feedback
between the experiment and theory during the exploratory search
for resonances. As the ABM model in its present form does not
yield the width of the resonances, the prediction of a resonance
position is not expected to be more accurate than the
corresponding experimental resonance width. For the
$^{6}$Li-$^{40}$K mixture, the ABM model predicts hundreds of
further resonances in various $s$- and $p$-wave channels up to
0.1\,T \cite{moredetails}.

A remarkable feature of the $^{6}$Li-$^{40}$K system is the large
widths of the $p$-wave resonances near 25\,mT, which by far
exceeds the width of the observed $s$-wave resonances. Naively,
one would expect the $s$-wave resonances to be wider than their
$p$-wave counterparts because of the different threshold behavior.
However, in the present case the difference in magnetic moments
between the atomic threshold and the relevant molecular state is
found to be anomalously small, which stretches out the thermally
broadened $p$-wave resonance features over an unusually wide
magnetic field range. Also the asymmetry of the loss feature
supports its interpretation as a $p$-wave resonance
\cite{Zhang2004pwfetal,Schunck2005frietal,Chevy2005rspetal}.

An important issue for future experiments is the character of the
$s$-wave resonances, i.e.\ the question whether they are
entrance-channel or closed-channel dominated
\cite{Kohler2006poc,Chin2008fri}. All our observed resonances are
rather narrow and thus closed-channel dominated. The existence of
entrance-channel dominated resonances would be of great interest
to experimentally explore BEC-BCS crossover physics
\cite{Inguscio2006ufg,Giorgini2007tou} in mixed Fermi systems.
However, our coupled channels calculations for a partial set of
predicted resonances have not yet found any such resonances, and
their existence seems unlikely in view of the moderate values of
the background scattering lengths
\cite{Chin2008fri,Kohler2006poc}.

In conclusion, we have characterized the interaction properties in
an ultracold mixture of $^6$Li and $^{40}$K atoms by means of
Feshbach spectroscopy and two theoretical models. The results are
of fundamental importance for all further experiments in the
emerging field of Fermi-Fermi mixtures. Further steps will be the
formation of bosonic $^6$Li$^{40}$K molecules through a Feshbach
resonance and evaporative cooling towards the creation of a
heteronuclear molecular Bose-Einstein condensate.

A double-degenerate mixture of $^6$Li and $^{40}$K was recently
demonstrated in a magnetic trap \cite{Taglieber2008qdtetal}.

\begin{acknowledgments}
We thank E.\ Tiemann for stimulating discussions. The Innsbruck
team acknowledges support by the Austrian Science Fund (FWF) and
the European Science Foundation (ESF) within the EuroQUAM project.
TGT and JTMW acknowledge support by the FOM-Program for Quantum
gases. SJJMFK acknowledges support from the Netherlands
Organization for Scientific Research (NWO). PSJ acknowledges
partial support by the U.S. Office of Naval Research.
\end{acknowledgments}

\bibliographystyle{apsrev}


\begin{thebibliography}{29}
\bibitem[{mor()}]{moredetails}
\bibinfo{note}{More details will be published elsewhere.}
\bibitem[{end()}]{endnoteOverlap}
\bibinfo{note}{The actual values are 0.979 for l=0 and 0.965 for l=1.}
\end{thebibliography}

\end{document}